\documentclass[pdftex,twocolumn,10pt,letterpaper]{article}
\usepackage{graphicx, times}
\usepackage{lipsum}
\usepackage{multirow}
\usepackage[utf8]{inputenc}
\usepackage[english]{babel}
\usepackage{float}
\usepackage{hyperref}
\usepackage{authblk}

\hypersetup{
    colorlinks=false,
    linkcolor=black,
    filecolor=black,      
    urlcolor=black,
}
 
\begin{document}
\title{Query Time Optimized Deep Learning Based Video Inference System}
\author[$\dagger$]{Mingren Shen}
\author[$\dagger$]{Shuoxuan Dong}
 \author[$\dagger$]{Xiuyuan He}

\affil[$\dagger$]{Department of Computer Sciences, University of Wisconsin - Madison, 1210 W Dayton St, Madison, WI 53706}

\interfootnotelinepenalty=10000

\maketitle

\begin{abstract}
	This is a project report about how we tune Focus\cite{hsieh2018focus}, a video inference system that provides low cost and low latency, through two phases. In this report, we will decrease the query time by saving the middle layer output of the neural network. This is a trade-off strategy that involves using more space to save time. We show how this scheme works using prototype systems, and it saves 20\% of the time. The code repository URL is here, \url{https://github.com/iphyer/CS744_FocousIngestOpt}.

\end{abstract}

\section{Introduction}
Video inference system has received increasing attention due to its potential in various real-world applications. By querying a video data set, we can answer questions like which frames contain objects of certain classes, such as certain types of cars, pedestrians, and bicycles. Answering these questions are very useful in different areas like traffic control, video surveillance, etc~\cite{hsieh2018focus}. 

Some may argue that video inference can be done using the same approach for image recognition, however, many existing  image recognition methods are too expensive to be  directly applied to video inference tasks. Even with state-of-the-art Convolutional Neural Networks (CNN) such as the Faster R-CNN~\cite{ren2017faster} running on powerful GPUs the speed is still relatively low and computational cost is high. According to literature~\cite{hsieh2018focus}, Faster R-CNN runs 30-80 fps on a \$ 4000 GPU and it can cost up to \$ 250 to query a month-long traffic camera video on cloud. This may seem to be affordable for videos from single camera, but imagine if you have videos from thousands of cameras need to be queried, the cost is will be prohibitively high. 

One of the cutting-edge video inference system that provides low cost and low latency is called Focus\cite{hsieh2018focus}. It uses an unique architecture that divide the video query problem into two phases: ingesting and querying. A cheap CNN is used in the ingesting phase to quickly generate Top-K indexes which contains K possible classes of the image. An expensive CNN is then used to generate a final and more accurate prediction during query time. 

\begin{figure*}[h]
    \centering
    \includegraphics[width=\textwidth]{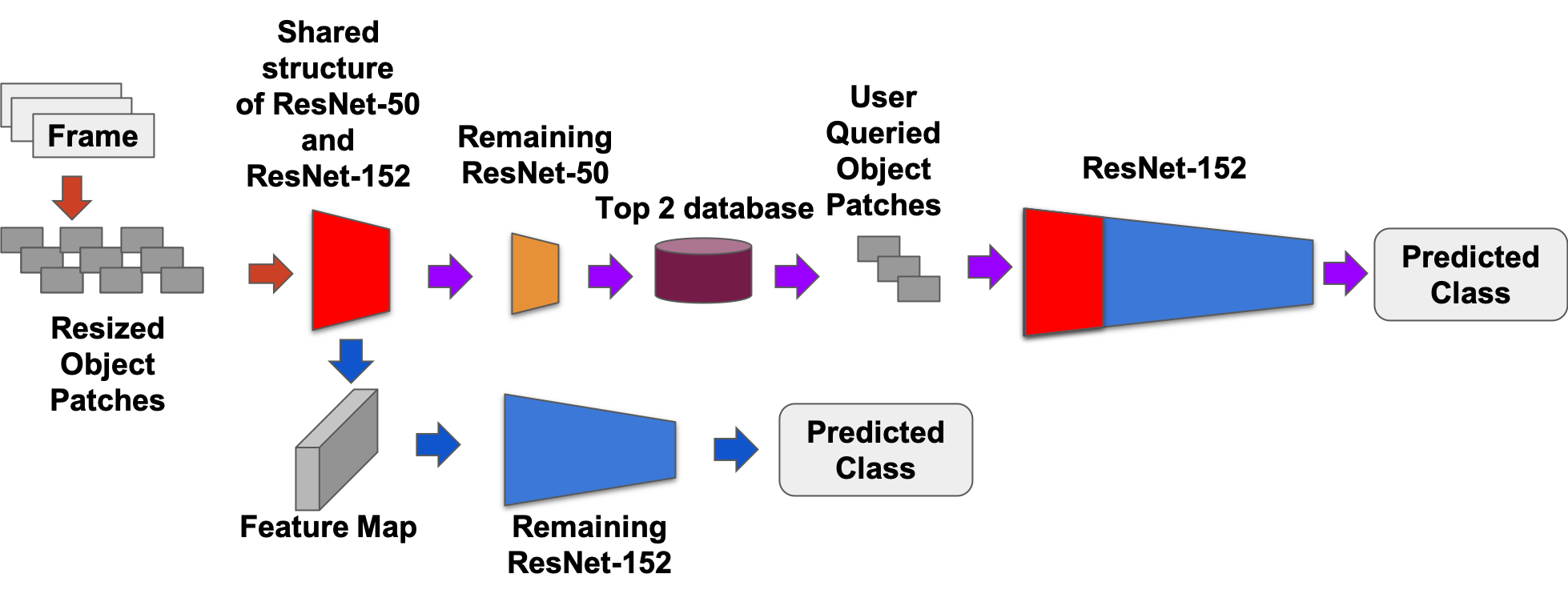}
    \caption{Project System Design}
    \label{fig:design}
\end{figure*}

For this project, we build our system based on the Focus architecture as shown in Fig~\ref{fig:design}. We propose to reuse the intermediate results of the shared layers(red trapezoid in Fig~\ref{fig:design}) between the cheap and expensive CNN to reduce inference time. Sharing convolution layers between two neural networks has already been used in Faster R-CNN model to speed up object detection where two networks, region proposal neutral network and Fast R-CNN~\cite{girshick2015fast}, share the beginning several convolution layers~\cite{ren2015faster,ren2017faster}.

In this report, We show how we use two CNNs, ResNet-50 and ResNet-152, to represent the architecture of Focus~\cite{hsieh2018focus} and we evaluate the accuracy and latency of it as our baseline. We then develop two different layer-sharing strategies. In one case the two networks share the first 4 blocks with each other, and in another case they share 7 blocks. The results suggest that depends on how many layers the two CNNs share, there is a 10\% to 20\% improvement for inference latency, as shown in Table~\ref{table:performance}. 

\section{Background}

\subsection{Video Inference System}
\label{sect:foucs}
Video inference system is an important area of computer vision research and application. Video inference systems often try to answer ``after-the-fact'' queries such as : ``In all the recent 24 hours surveillance videos from 1st ave to 5th ave, return all the frames containing a truck." Such types of questions are frequently asked by local police and other agencies~\cite{hsieh2018focus}. NoScope and Focus are two state-of-the-art systems designed to answer these questions. This report is mainly based on Focus.

As mentioned above, Focus has two phases: ingesting and querying. During ingesting time, Focus uses a cheap CNN model, ResNet-18~\cite{he2016deep}, to pick the Top-K Indexes for each frame. For example, if the original image is a truck, the index can be a list of classes like [boat, car, truck, bus]. The first prediction on that list may not be the most accurate one. However, the top-K list almost  always contain the class “truck”. During querying time, a user can query for a certain class like ``truck". With the Top-K Indexes from ingesting time, the system identifies the frames that are possibly interesting to the query. An expensive CNN model, ResNet-152, is then used to do predictions only on the selected frame set, whose index contains the specified class user queried. Such an expensive model  can cost 8$\times$ more compared to its cheaper variant ResNet-18, but can yield more accurate results. 

\subsection{Convolutional Neural Networks}
Convolutional Neural Networks(CNNs) are a special class of neural network that use many layers of convolution filters to automatically extract hierarchical helpful features for the final task like classification. For example, for image classification task, beginning layers of CNNs try to learn lower-level features such as dots, lines and edges. Layers in the middle tend to build object parts based on those lower-level features and layers near the output are learning the most semantic and high-level object related information. Finally, a score is computed for each image class as output~\cite{lecun2015deep}. 

Since Krizhevsky et al. won ImageNet competition in 2012~\cite{krizhevsky2012imagenet}, CNNs are the most popular methods for image related tasks~\cite{lecun2015deep}. In 2015, a new deep residual learning method called ResNet~\cite{he2016deep} even surpassed human performance on image classification tasks of ImageNet. ResNet contains many stacked basic units called blocks to solve the degradation problem encountered in training deep neural network~\cite{he2016deep}. The central idea is that by attaching an identity mapping with skip connections, even very deep layers in ResNet can get the identical input from previous layer and the newly added layer should not perform worse than the shallower model~\cite{he2016deep}.

CNNs are not only useful for image related tasks, they can also be used for videos~\cite{hsieh2018focus,redmon2018yolov3,kang2017noscope} inference and tracking,  recommended systems~\cite{van2013deep}, natural language processing~\cite{collobert2008unified}, and physics~\cite{baldi2014searching} ,chemistry~\cite{segler2018planning,goh2017deep} , material sciences~\cite{sun2020assessing,jacobs2022performance,field2021development,field2020rapid,shen2022machine,morganautomated,shen2021deep,shen2021machine, shen2021multi,lawrence2020exploring,morganautomated, jacobs2021performance}  or medical image~\cite{liu2019harmonization, gurbani2021evaluation,awe2022machine} and other problems~\cite{shen2019diffusive,ming2016mesoscale,shen2016chemically,luo2018n6, sun2020assessing}.

\subsection{Traffic Video Characteristic}
One important characteristic of surveillance or traffic camera videos is that large portion of the frames from the video is duplicate and useless, for example, if the scene is nearly static. So processing each frame is less efficient. To address this problem, NoScope~\cite{kang2017noscope} uses a difference detector that measures the Mean Square Error (MSE) between images to determine whether the content of the frame has changed. Compared with NoScope, Focus uses a different approach to solve the duplicate frame problem. It clusters the frames based on the output of the penultimate (i.e., previous-to-last) layer~\cite{hsieh2018focus}.

We create our data-set based on the traffic camera video frames from Urban Tracker Dataset~\cite{Jodoin2014UrbanTracker}. The patches containing objects are extracted from the original images using background subtraction. The patches are labeled with 3 classes including pedestrians, cars and some subtracted background. The models are trained with manually labeled image patches.

\subsection{Feature Map }
Feature Maps or Feature Vectors is the intermediate output from certain CNN layers. They are abstract visual features and high-level representation of the original images. Feature maps can gradually capture higher level features that consist of lower level features. Caching feature maps in memory or save them to disk can save inference time for further uses. Xu et al. developed an effective CNN cache system to store these feature maps and reuse them at fine spatial granularity to get high performance when doing inference in continuous mobile vision~\cite{xu2017deepcache}. Indeed, the data volume of feature maps is large, so there might be a trade off between efficiency(inference time) and storage (memory) to reuse feature maps.

\section{Design}
\subsection{Dataset Generation}
\begin{figure}[h]
    \centering
    \includegraphics[width=0.5\textwidth]{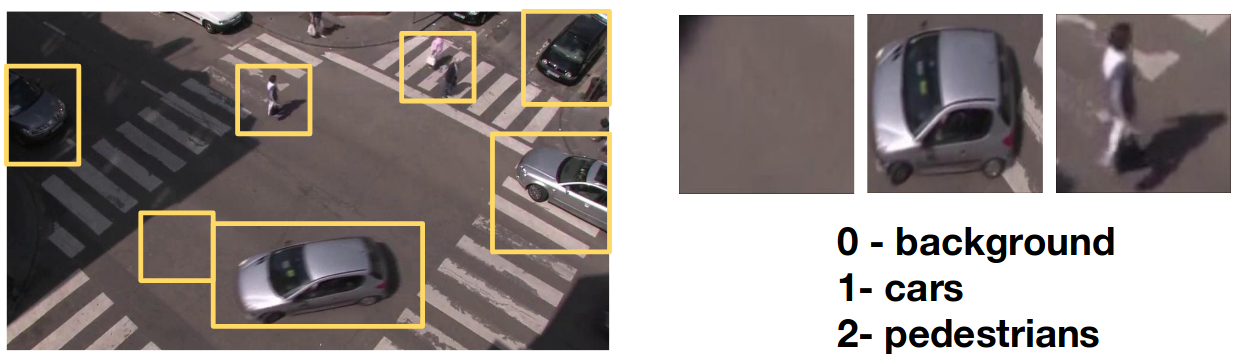}
    \caption{Background Subtraction and Generated Patches}
    \label{fig:bgs}
\end{figure}

A video from the Urban Tracker Dataset~\cite{Jodoin2014UrbanTracker} is used to generate object patches. To convert video into object patches, we use background subtraction to detect the moving objects (cars, pedestrians, etc.). The detector we use is BackgroundSubtractorGMG in OpenCV, which is an algorithm that combines per-pixel Bayesian segmentation with statistical background image estimation~\cite{godbehere2012visual}. We generate 2275 patches with 3 classes: cars, pedestrians and background images (noise generated by background sub-tractor), as shown in Fig~\ref{fig:bgs}. All patches are re-scaled to 224x224 pixels.

\subsection{Building Baseline}
We jointly use two networks to mimic the two CNNs architecture of Focus~\cite{hsieh2018focus}. Like Focus, we use a cheap CNN for ingest phase, during which top-k indexes are generated. During query time, the user queries frames containing a specified class. An expensive CNN is used to classify the images only if the top-k index of the image contains the desired object class.

\subsection{Feature Map Reuse}
\begin{figure*}[h]
    \centering
    \includegraphics[width=\textwidth]{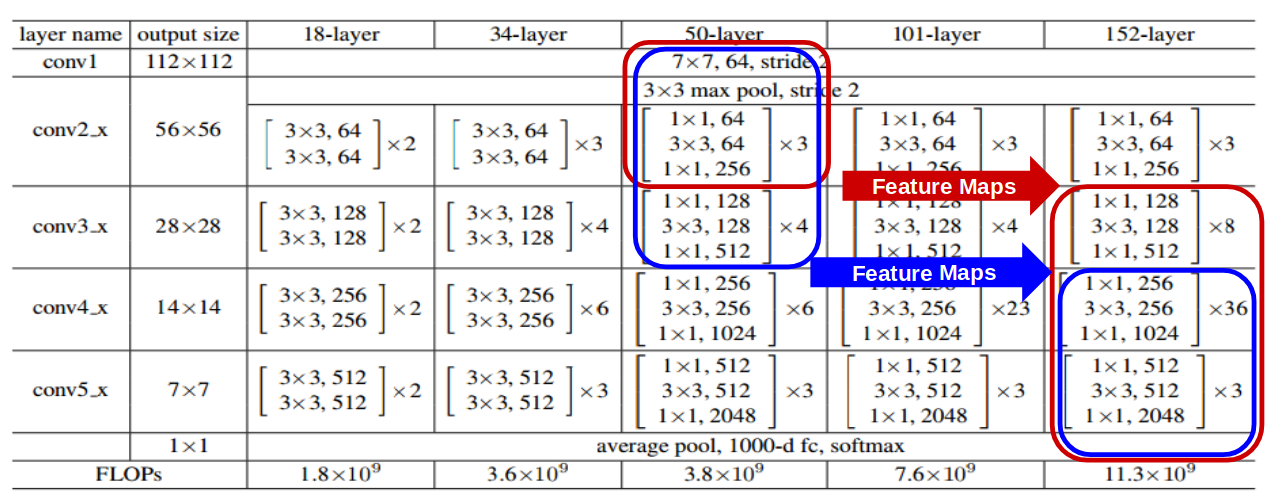}
    \caption{ResNet architecture and two different cuts.The red color reuses 4 blocks and the blue one reuses 7 blocks. }
    \label{fig:resnetarch}
\end{figure*}

Focus uses ResNet-18 and 152 as the cheap and expensive CNN, whose structure is shown in Fig~\ref{fig:resnetarch}. From that figure, we can find that the two CNNs have some same structure layers since the input layer. Considering there are two different residual blocks in ResNet~\cite{he2016deep}, the basic block and the bottleneck block, we only reuse shared layers if the two models are built upon the same residual blocks. For example, if comparing ResNet-18 and ResNet-34, the two model are both built with basic blocks, and the first two residual blocks have the exact same structure. Another example would be ResNet-50 and ResNet-152, both are built with bottleneck blocks and they share the same structure for the first seven blocks. By reusing the shared structure of two networks, we try to reduce computation amount during the query phase.

\subsection{Overall Experiment Design}
We run our experiment based on ResNet-50 and 152, in this way we can skip more blocks, compared with ResNet-18 and 34.

During ingest time, we classify all the patches using ResNet-50 and generates Top-3 indexes. At the same time, we save the feature map for the first several residual blocks and reuse it as the input of the expensive CNN.  Here we tested two cases: one is reusing the first 4 blocks and another is reusing the first 7 blocks. For the query phase, we run the expensive model to predict on the feature map. We evaluate the performance with and without retrain the remaining layers of the model.

\section{Evaluation}

A ResNet-50 model and a ResNet-152 are trained on the data-set we mentioned above\footnote{ Our system is built on the basis of a Keras-ResNet repository on Github \url{https://github.com/raghakot/keras-resnet}}. The ratio of training set and testing set is 4:1. We also retrain the remaining part of ResNet-152 with the feature maps that are output from the shared layers of the trained ResNet-50. All training curves are shown in the Appendix. The models are able to achieve almost 1 on training data-set and 0.95 to 0.97 on the testing data-set.

The developed system with trained model parameters is tested on a laptop computer to evaluate the latency of inference\footnote{Dell Inspiron 7000 laptop with a i7-6700hq CPU and 16 GBs of memory.}. The performance results of the inference system are shown in table~\ref{table:performance}. For each scenario, the test is done 5 times to show average and standard deviation of latency.

It is observed that the feature map reusing models with weights trained from original structures cause huge loss in accuracy, which means that the feature maps from the shared layers of ResNet-50 model can not be fully understood by ResNet-152 although they have the same dimension. So, retrain the remaining layers of ResNet-152 would be the proper way to build the system without accuracy sacrifice.

We are able to achieve a 9.6\% reduction of query time for the shallow cut retrained model with a compromise of 0.9\% accuracy. Moreover, we get 17.5\% reduction of query time for the deep cut retrained model with no accuracy reduction.

\begin{table*}[h!]
\centering
\begin{tabular}{||c|c|c|c|c|c||} 
 \hline
 \hline
 Model & Accuracy & Accuracy & Latency(sec/image) & Latency & Latency \\ 
 & & Reduction & & Standard Deviation & Reduction\\
 \hline\hline
 ResNet-50 + 152 &
 \multirow{2}{2em}{0.967} &
 \multirow{2}{1em}{N/A} &
  \multirow{2}{2em}{0.3611} &
 \multirow{2}{1em}{0.0038} & \multirow{2}{1em}{N/A} \\
 (Focus Baseline)  &&&&&\\
 \hline
 Reuse Feature Map & 
 \multirow{2}{2em}{0.518} & \multirow{2}{1em}{46.4\%} &
  \multirow{2}{2em}{0.3277} & \multirow{2}{1em}{0.0072} & \multirow{2}{1em}{10.6\%} \\
(shallow cut)  &&&&&\\
 \hline
 Reuse Feature Map  &  
 \multirow{2}{2em}{0.958} & \multirow{2}{1em}{0.9\%} &  \multirow{2}{2em}{0.3266}& \multirow{2}{1em}{0.0069} &  \multirow{2}{1em}{9.6\%} \\ 
 (shallow cut, retrained) &&&&&\\
 \hline
 Reuse Feature Map  &  
 \multirow{2}{2em}{0.391} & \multirow{2}{1em}{59.6\%} &  \multirow{2}{2em}{0.3058} & \multirow{2}{1em}{0.0078} & \multirow{2}{1em}{15.3\%} \\
 (deep cut) &&&&&\\
 \hline
 Reuse Feature Map  &  
 \multirow{2}{2em}{0.969} &
 \multirow{2}{1em}{0\%} &  \multirow{2}{2em}{0.2978} & \multirow{2}{1em}{0.0070} & \multirow{2}{1em}{17.5\%} \\ 
 (deep cut, retrained) &&&&&\\
 \hline
\end{tabular}
\caption{System performance of difference cuts}
\label{table:performance}
\end{table*}

\section{Related Work}

\subsection{Object Detection}
Most existing object detection techniques, including two-stage and single-stage methods, rely on Convolution Neural Networks~\cite{liu2018deep}. A two-stage method is a region proposal based method. It first generates region proposals and then predicts bounding boxes based on these proposed regions. Many researchers have been focusing on improving the performance and speed of two-stage method like R-CNN~\cite{girshick2014rich}, Fast R-CNN~\cite{girshick2015fast}, Faster R-CNN~\cite{ren2015faster,ren2017faster}, etc. Compared to the two-stage method, the single-stage method is usually faster and simpler~\cite{liu2018deep}. So, the single-stage method is popular in speed demanding object detection tasks. For example, YOLOv1~\cite{redmon2016you} could perform real-time object detection in videos when most two-stage methods fail.

\subsection{Other Video Inference Systems}
Besides Focus, NoScope is another system for accelerating neural network query speed over videos~\cite{kang2017noscope}. To reduce the cost of doing Neural Network inference, NoScope uses a specialized shallower model that is less generalized but faster. The specialized model estimates the class of the objects in the frame with a confidence score. If the confidence is not high enough, a more accurate but higher cost model will be called to generate the final class for this object. However, NoScope is limited since it relies binary classifiers which is hard to scale.

\section{Future work}
First, our system needs to be tested on other datasets with more images and more classes. The current dataset contains just small patches of 3 different classes with a low resolution. Testing on more higher resolution datasets can prove the robustness of the system. We would also like to explore strategies of reusing layers on different ResNet models, e.g. combine ResNet-18 and 152, on the inference time. A generic methodology for deciding where to cut the layers would be useful which may need the statistics of the time consumed on each layer. Moreover, we currently only focus on ResNet architecture, which imposes restrictions for the application of this system. Other types of models should also be considered. 

Since we are building our system on top of Focus~\cite{hsieh2018focus}. The other features in Focus should be realized in our system and some of these features can even be optimized. For example, the clustering of objects in Foucs can help reduce duplication of object patches, only uses simple incremental clustering method which is well-studied in~\cite{hsieh2018focus,o2002streaming,cao2006density}. However, this approach has some drawbacks. First, it relies on the hyper-parameters manually set before running the algorithm. Second, the clustering parameter T, the distance threshold between clusters, has a direct impact on both recall and precision. This requires the user has background knowledge  about how to configure the clustering algorithm. Also, more clustering algorithms can be tested for better performance.

{
\bibliographystyle{unsrt}
\bibliography{ref}
}
\section{Appendix}
\subsection{Training Curves}

\begin{figure}[h]
    \centering
    \includegraphics[width=0.3\textwidth]{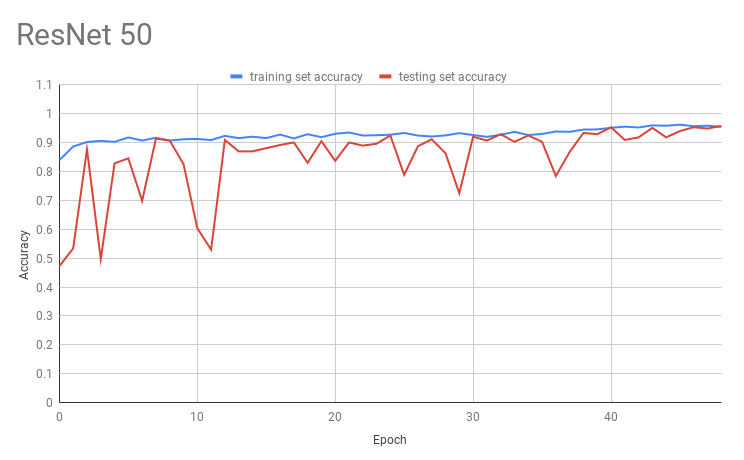}
    \caption{Training Curves of ResNet-50}
    \label{fig:curve50}
\end{figure}

\begin{figure}[h]
    \centering
    \includegraphics[width=0.3\textwidth]{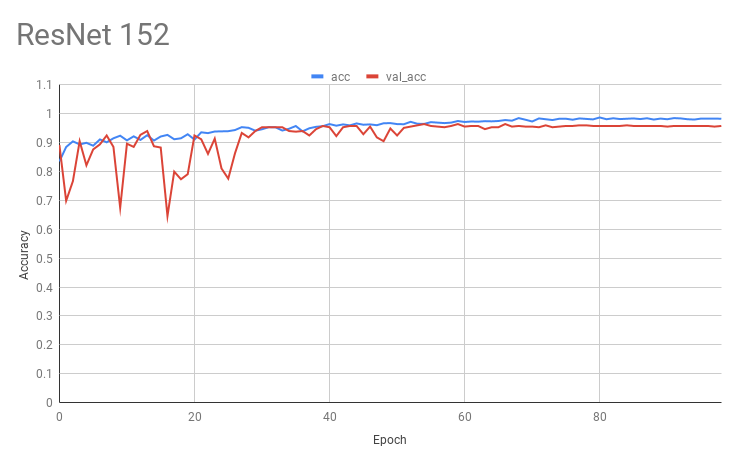}
    \caption{Training Curves of ResNet-152}
    \label{fig:curve152}
\end{figure}

\begin{figure}[h]
    \centering
    \includegraphics[width=0.3\textwidth]{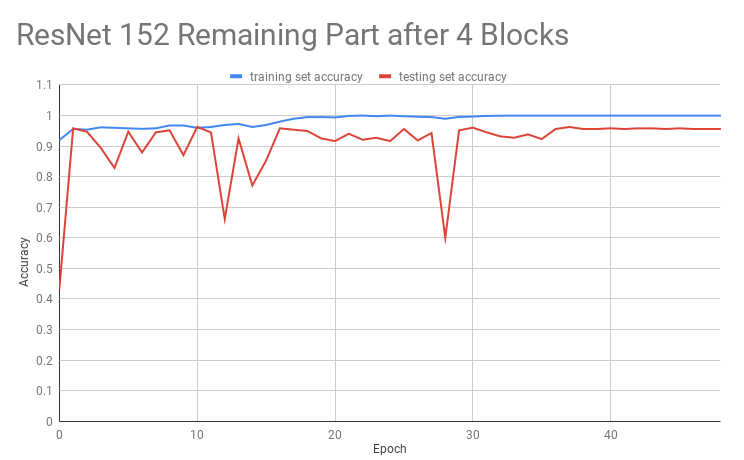}
    \caption{Training Curves of ResNet-152 cutting 4 blocks}
    \label{fig:curve152-4}
\end{figure}

\begin{figure}[h]
    \centering
    \includegraphics[width=0.3\textwidth]{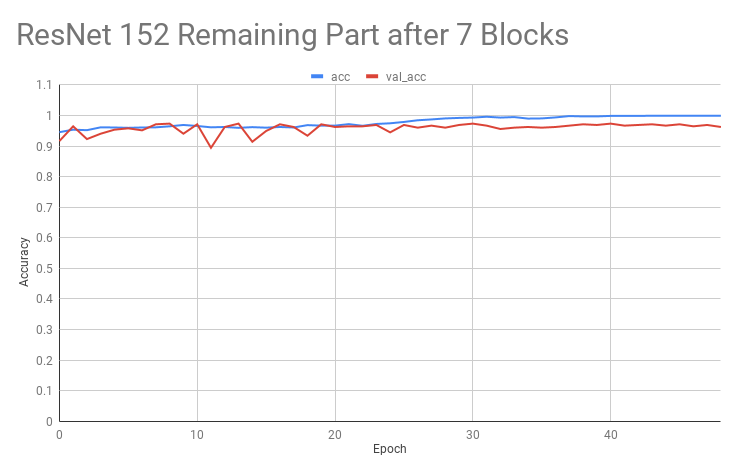}
    \caption{Training Curves of ResNet-152 cutting 7 blocks}
    \label{fig:curve152-7}
\end{figure}

The models converge after about 50 to 100 learning epochs and the accuracy of evaluation comes to or above 95\%.
The learning curves for ResNet-50, ResNet-152 and the remaining part of ResNet-152 after two cuts is shown in Figure ~\ref{fig:curve50}, Figure ~\ref{fig:curve152}, Figure ~\ref{fig:curve152-4}, Figure ~\ref{fig:curve152-7}.


\subsection{Source Code of this Project}

We provide all the source code of this project on Github and you check the code from this link,\url{https://github.com/iphyer/FocousIngestOpt_FinalProject_CS744Fall2018}. 

The code is built on Python3 and Keras with Tensorflow as back-end engine.
\end{document}